\documentclass[manuscript,screen]{acmart}
\usepackage{listings}
\usepackage{tcolorbox}
\AtBeginDocument{%
  }

\setcopyright{acmlicensed}
\copyrightyear{2025}
\acmYear{2025}
\acmDOI{XXXXXXX.XXXXXXX}

\acmISBN{978-1-4503-XXXX-X/2018/06}

\begin{document}
\title{Interacting with AI Reasoning Models:\\Harnessing ``Thoughts'' for AI-Driven Software Engineering}

\author{Christoph Treude}
\affiliation{%
  \institution{Singapore Management University}
  \city{Singapore}
  \country{Singapore}}
\email{ctreude@smu.edu.sg}

\author{Raula Gaikovina Kula}
\affiliation{%
  \institution{Osaka University}
  \city{Osaka}
  \country{Japan}}
\email{raula-k@ist.osaka-u.ac.jp}

\begin{abstract}
Recent advances in AI reasoning models provide unprecedented transparency into their decision-making processes, transforming them from traditional black-box systems into models that articulate step-by-step chains of thought rather than producing opaque outputs. This shift has the potential to improve software quality, explainability, and trust in AI-augmented development. However, software engineers rarely have the time or cognitive bandwidth to analyze, verify, and interpret every AI-generated thought in detail. Without an effective interface, this transparency could become a burden rather than a benefit.

In this paper, we propose a vision for structuring the interaction between AI reasoning models and software engineers to maximize trust, efficiency, and decision-making power. We argue that simply exposing AI’s reasoning is not enough -- software engineers need tools and frameworks that selectively highlight critical insights, filter out noise, and facilitate rapid validation of key assumptions. To illustrate this challenge, we present motivating examples in which AI reasoning models state their assumptions when deciding which external library to use and produce divergent reasoning paths and recommendations about security vulnerabilities, highlighting the need for an interface that prioritizes actionable insights while managing uncertainty and resolving conflicts.
We then outline a research roadmap for integrating automated summarization, assumption validation, and multi-model conflict resolution into software engineering workflows. Achieving this vision will unlock the full potential of AI reasoning models to enable software engineers to make faster, more informed decisions without being overwhelmed by unnecessary detail.
\end{abstract}




\maketitle

\section{Introduction}

Recent advances in AI reasoning models have led to higher quality outputs because these models engage in structured reasoning rather than merely generating responses~\cite{guo2025deepseek}. By explicitly working through logical steps, AI reasoning models have the potential to produce more accurate, insightful, and useful recommendations for software engineers~\cite{li2025structured}. However, beyond just improving output quality, these models also help address the long-standing problem of AI being a black box~\cite{cao2024systematic}. Instead of producing results without explanation, they now reveal their thought processes, making it possible to understand why AI reaches specific conclusions and how it evaluates alternatives~\cite{wangunlocking}.

This ability to ``show their thoughts'' represents a fundamental shift in AI-assisted software development. It enables developers not only to trust AI-generated recommendations, but also to verify and scrutinize the reasoning behind them. This shift means that developers are no longer forced to rely on AI without knowing why it works or how it comes to decisions~\cite{drozdal2020trust}. By making AI's thought process transparent, these models help overcome the historical challenge of opaque AI decision making~\cite{dehghani2024trustworthy} and allow developers to critically assess each step before taking action.

Despite these improvements, the sheer volume of AI-generated reasoning paths can be overwhelming. Developers may be confronted with redundant details, implicit assumptions, and conflicting conclusions, making it difficult to effectively extract actionable insights~\cite{terragni2025future}. Without methods to interpret and validate these thought traces, AI reasoning can become more of a burden than a benefit.

To address these challenges, we propose a vision for structured interaction with AI reasoning models. Instead of presenting raw thought traces, AI systems should filter, summarize, and highlight key decision factors, ensuring that developers can efficiently validate assumptions and compare alternative reasoning paths. We illustrate these issues through two motivating examples in which AI-generated reasoning significantly influences software decisions: one where implicit assumptions lead to a library choice recommended by AI that may not align with the needs of the developer, and another where different AI reasoning models provide conflicting perspectives on security vulnerabilities.
This paper outlines a research roadmap for harnessing AI reasoning in software engineering, focusing on automated summarization, assumption extraction, and multi-model alignment. Our goal is to enable software engineers to benefit from the power of AI-driven reasoning without being overwhelmed by unnecessary detail, ultimately improving trust, efficiency, and decision making in AI-assisted software development.

\section{Motivating Examples}

In this section, we present two examples in which being aware of AI reasoning would have the potential to significantly impact software engineering decisions. The complete prompts, reasoning paths, and responses for these examples are available in our online appendix at \url{https://doi.org/10.5281/zenodo.14946531}.

\subsection{Hidden Assumptions in AI Reasoning}

\begin{figure}[t]
    \centering
\begin{tcolorbox}[
    title=Motivating Example 1: Generating a PDF Report, 
    fonttitle=\bfseries, 
    coltitle=white, 
    colback=white, 
    colframe=black, 
    colbacktitle=black,
    sharp corners=south
]

\noindent \textbf{Prompt:} Modify this function to generate a PDF report from the given data.

\begin{lstlisting}[language=Python]
def generate_report(data):
    report_content = f"Financial Summary:\n{data}"
    return report_content
\end{lstlisting}

\begin{center}
    \tcbsubtitle[before skip=0pt, after skip=10pt, colback=black, coltitle=white, fonttitle=\bfseries]{Thought Process of Gemini 2.0 Flash Thinking Experimental 01-21}
\end{center}

\noindent \textbf{Here's a breakdown of my thought process in approaching the request to generate a PDF report from data:}

\begin{enumerate}
    \item \textbf{Understanding the Core Requirement:} 
    The user wants to take some \texttt{data} (presumably financial data, based on the "Financial Summary" string) and turn it into a PDF report. The original function just returned a plain text string. The key is to move from plain text to a formatted PDF document.

    \item \textbf{Identifying the Need for a PDF Library:} 
    Python doesn't have built-in PDF generation capabilities. We need a library. Several options come to mind:
    \begin{itemize}
        \item \textbf{ReportLab:} Powerful, feature-rich, but can have a steeper learning curve. Excellent for complex layouts and professional reports.
        \item \textbf{PyFPDF:} Simpler to use, good for basic text and images in PDFs. More lightweight than ReportLab.
        \item \textbf{xhtml2pdf (or WeasyPrint):} Converts HTML/CSS to PDF. Excellent if you're comfortable with web technologies and want more visually appealing layouts.
        \item \textbf{Other libraries (like PDFKit, which wraps wkhtmltopdf):} Also HTML-to-PDF based.
    \end{itemize}

    For this example, since the initial request is simple (just "Financial Summary" and data), and for ease of demonstration, \textbf{PyFPDF} seems like a good starting point. It's relatively easy to install and use for basic text-based PDFs. If the user needed more complex layouts, images, tables, etc., I might have considered ReportLab or an HTML-to-PDF approach.

\end{enumerate}
[...]
\end{tcolorbox}
    \caption{First Motivating Example}
    \label{fig:motivating-example-1}
\end{figure}

Decision making in software engineering often relies on implicit assumptions. In the first example shown in Figure~\ref{fig:motivating-example-1}, Google's Gemini Flash Thinking~\cite{deepmindGeminiFlash} is asked to modify a function to generate a PDF report. The AI correctly identifies that an external library is needed and recommends PyFPDF. Looking at the reasoning of the model, it turns out that this recommendation is based on an assumption made by the model: ``For this example, since the initial request is simple (just "Financial Summary" and data), and for ease of demonstration, PyFPDF seems like a good starting point.''

If the developer requires advanced formatting, embedded images, or interactive elements, PyFPDF may be the wrong choice. A more suitable library, such as ReportLab or an HTML-to-PDF tool, might be needed. Another AI reasoning model, OpenAI o1~\cite{jaech2024openai}, indeed suggests Foxit PDF, highlighting how different reasoning paths lead to different conclusions. Without visibility into AI assumptions, developers risk blindly following recommendations that may not fit their actual needs.

\subsection{Diverging AI Reasoning Paths}

\begin{figure}[t]
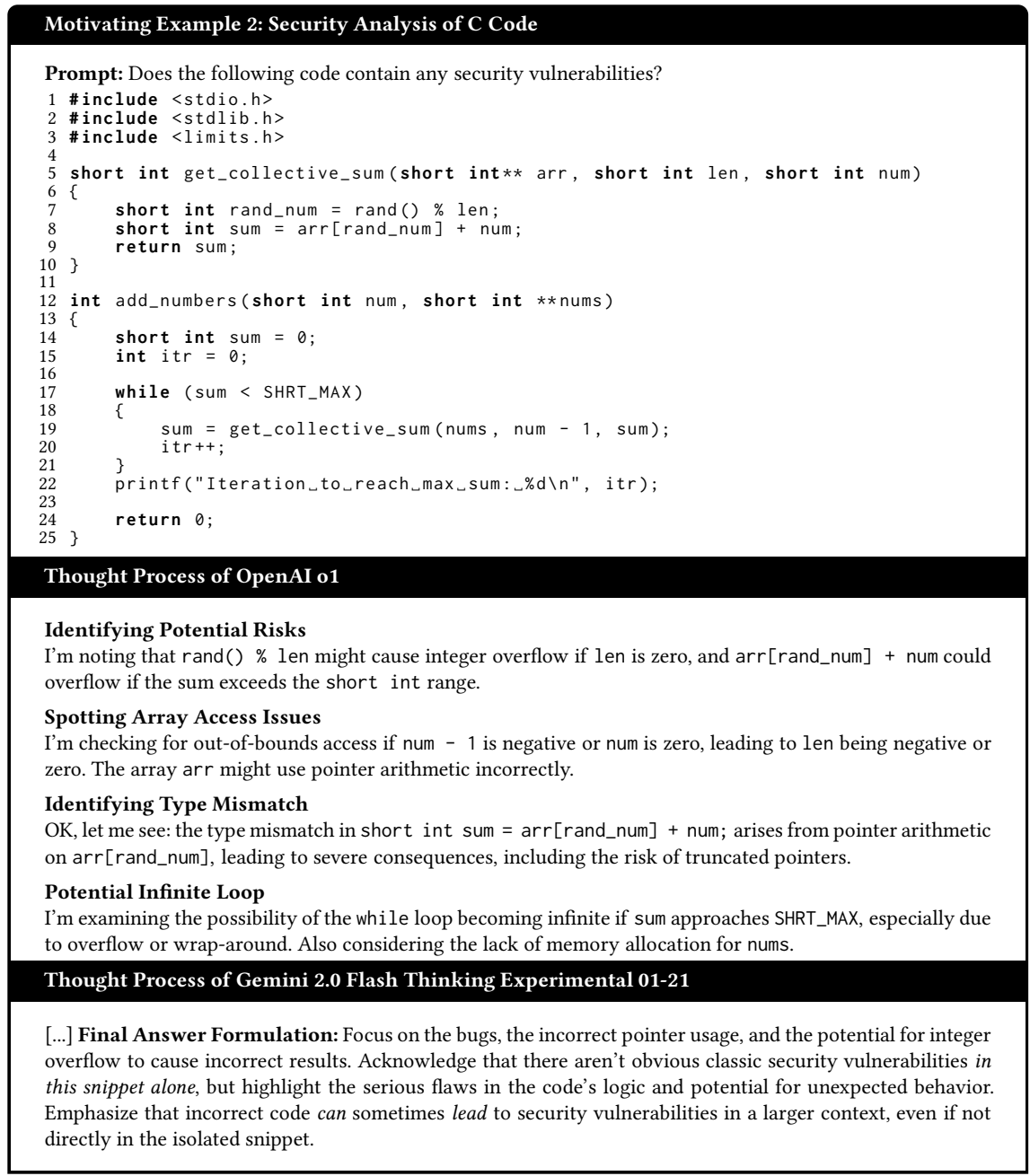

    \centering
\begin{tcolorbox}[
    title=Motivating Example 2: Security Analysis of C Code, 
    fonttitle=\bfseries, 
    coltitle=white, 
    colback=white, 
    colframe=black, 
    colbacktitle=black,
    sharp corners=south
]

\noindent \textbf{Prompt:} Does the following code contain any security vulnerabilities?

\begin{lstlisting}[language=C]
#include <stdio.h>
#include <stdlib.h>
#include <limits.h>

short int get_collective_sum(short int** arr, short int len, short int num)
{
    short int rand_num = rand() % len;
    short int sum = arr[rand_num] + num;
    return sum;
}

int add_numbers(short int num, short int **nums)
{
    short int sum = 0;
    int itr = 0;
    
    while (sum < SHRT_MAX)
    {
        sum = get_collective_sum(nums, num - 1, sum);
        itr++;
    }
    printf("Iteration to reach max sum: %d\n", itr);

    return 0;
}
\end{lstlisting}

\begin{center}
    \tcbsubtitle[before skip=0pt, after skip=10pt, colback=black, coltitle=white, fonttitle=\bfseries]{Thought Process of OpenAI o1}
\end{center}

\noindent \textbf{Identifying Potential Risks}

I'm noting that \texttt{rand() \% len} might cause integer overflow if \texttt{len} is zero, and \texttt{arr[rand\_num] + num} could overflow if the sum exceeds the \texttt{short int} range.

\medskip

\noindent \textbf{Spotting Array Access Issues}

I'm checking for out-of-bounds access if \texttt{num - 1} is negative or \texttt{num} is zero, leading to \texttt{len} being negative or zero. The array \texttt{arr} might use pointer arithmetic incorrectly.

\medskip

\noindent \textbf{Identifying Type Mismatch}

OK, let me see: the type mismatch in \texttt{short int sum = arr[rand\_num] + num;} arises from pointer arithmetic on \texttt{arr[rand\_num]}, leading to severe consequences, including the risk of truncated pointers.

\medskip

\noindent \textbf{Potential Infinite Loop}

I'm examining the possibility of the \texttt{while} loop becoming infinite if \texttt{sum} approaches \texttt{SHRT\_MAX}, especially due to overflow or wrap-around. Also considering the lack of memory allocation for \texttt{nums}.

\begin{center}
    \tcbsubtitle[before skip=0pt, after skip=10pt, colback=black, coltitle=white, fonttitle=\bfseries]{Thought Process of Gemini 2.0 Flash Thinking Experimental 01-21}
\end{center}

\noindent {[...]} \textbf{Final Answer Formulation:} Focus on the bugs, the incorrect pointer usage, and the potential for integer overflow to cause incorrect results. Acknowledge that there aren't obvious classic security vulnerabilities \textit{in this snippet alone}, but highlight the serious flaws in the code's logic and potential for unexpected behavior. Emphasize that incorrect code \textit{can} sometimes \textit{lead} to security vulnerabilities in a larger context, even if not directly in the isolated snippet.

\end{tcolorbox}
    \caption{Second Motivating Example}
    \label{fig:motivating-example-2}
\end{figure}

The second example shown in Figure~\ref{fig:motivating-example-2} explores how different AI reasoning models use different reasoning paths when analyzing security vulnerabilities in C code. The code snippet was taken from related work~\cite{ullah2024llms} with the following ground truth: ``The function `add\_numbers` initializes a short int `sum` and in a while loop calls `get\_collective\_sum` to accumulate `sum` with a random number from the `nums` array. While doing so the `sum` can overflow and wraparound. Which can make the while loop run for an infinite time. This can cause an integer overflow attack.''. OpenAI o1 immediately focuses on security risks, identifying integer overflow, out-of-bounds memory access, and a potential infinite loop. By contrast, Gemini Flash Thinking appears to focus on general bugs rather than security vulnerabilities. It correctly identifies problematic code but does not explicitly connect these issues to security risks.

A developer who only looks at the final output of Gemini Flash Thinking would not realize that the AI got distracted by general software defects, as evident from its thought process. As a result, a developer might mistakenly assume that no security vulnerabilities exist when, in reality, the AI simply failed to recognize their implications.

\section{The Challenge: From Black-Box to Thoughtful Collaboration}

Reasoning has long played an important role in software engineering, particularly through symbolic reasoning, where logic-based systems formally verify software properties~\cite{yu2018symbolic}, check constraints~\cite{ben2015symbolic}, and find bugs~\cite{de2010bugs}. However, the rise of AI reasoning models introduces a new kind of reasoning -- one that is less about strict formalism and more about adaptive thinking processes based on natural language~\cite{mirzadeh2024gsm}. Unlike symbolic reasoning, which relies on explicit rules and proofs, AI reasoning models generate step-by-step explanations in free text, mixing observations, assumptions, and conclusions in ways that are often useful but difficult to verify~\cite{wei2022chain}.

Despite the lack of determinism, many software engineers have embraced AI, e.g., to generate code~\cite{liu2024your}, code reviews~\cite{lu2023llama}, bug reports~\cite{shi2022buglistener}, commits~\cite{zhang2024automatic}, Stack Overflow answers~\cite{zhong2024can}, and tests~\cite{schafer2023empirical}. This growing reliance on AI models and their reasoning raises questions about their usability and trustworthiness in software engineering.

One key challenge is information overload. AI-generated reasoning is often verbose, requiring developers to extract relevant insights manually. While symbolic reasoning produces structured, rule-based outputs, AI reasoning models generally lack formal guarantees~\cite{kamran2024vision}, making it harder to assess their reliability without human review. To benefit from transparent reasoning, developers must determine which parts of AI reasoning are correct, which are assumptions, and which are potentially misleading.

Second, conflicting reasoning paths~\cite{tang2023towards}: Different AI models, or even different runs of the same model, may produce divergent reasoning traces for the same problem. In contrast, symbolic reasoning is deterministic -- given the same inputs, it produces the same results~\cite{weng2023mastering}. AI reasoning models, however, may prioritize different aspects of a problem, for example, leading one to focus on security risks while another highlights performance concerns. This is particularly relevant in multi-agent environments~\cite{he2025llm}. Without a way to compare and reconcile these conflicting perspectives, software engineers may struggle to determine which reasoning to trust.

Third, trust and usability~\cite{wang2024investigating}: AI reasoning models provide more transparency than black-box predictions, but transparency alone does not guarantee usefulness. Developers must still verify whether the AI reasoning aligns with domain knowledge and best practices. Without an interface that filters out irrelevant details, highlights critical assumptions, and allows validation, the burden remains on developers to manually extract useful insights, making AI-generated reasoning as much a cognitive challenge as a technical one.

Unlike other fields where AI reasoning models are used primarily for classification and prediction, software engineering requires AI to engage in complex, context-sensitive reasoning that directly influences developers’ decision-making. Software engineers must assess AI-generated insights not only for accuracy, but also for relevance within architectural constraints, evolving project requirements, and team practices. Furthermore, software engineering is highly iterative and collaborative, meaning that AI-generated reasoning must align with human workflows, coding conventions, and best practices while balancing trade-offs in performance, maintainability, and security. Without interfaces that help software engineers efficiently interpret, validate, and integrate AI reasoning, the cognitive burden of interacting with these models may outweigh their benefits.

\section{Vision: A Human-AI Interface for Thoughtful Software Engineering}

Research on human-AI interfaces in software engineering is still in its infancy~\cite{treude2025developers}. To address these challenges, we need structured interfaces for interacting with AI-generated reasoning. These interfaces must summarize, filter, and compare AI thought processes, allowing developers to quickly assess assumptions, resolve conflicting reasoning paths, and focus on insights that truly impact their decision making. Unlike traditional symbolic reasoning systems, which are rule-based and deterministic, AI reasoning models are contextual and probabilistic, requiring a different approach to validation and integration into software engineering workflows.

\subsection{Filtering and Summarization}

AI reasoning models generate extensive thought traces, but not all information is equally useful. Some reasoning steps may be trivial, redundant, or overly verbose, while others contain critical insights that influence decision making. The interface should:
\begin{itemize}
    \item Distill long reasoning traces into concise, structured summaries that highlight key takeaways.
    \item Automatically detect and surface the most relevant reasoning steps while filtering out irrelevant details.
    \item Enable customizable views so that developers can choose between high-level summaries and detailed step-by-step reasoning when needed.
\end{itemize}

Summarization techniques, such as hierarchical reasoning trees or interactive explanations, could help developers engage with AI reasoning at different levels of detail, ensuring that transparency does not come at the cost of usability. Work in this area could build on a long line of work in software engineering on information filtering~\cite{kersten2006using} and summarization~\cite{haiduc2010use}.

\subsection{Assumption Extraction and Verification}

AI reasoning models frequently make implicit assumptions that shape their recommendations. If these assumptions remain hidden, developers may unknowingly accept recommendations that do not align with their actual requirements. The interface should:
\begin{itemize}
    \item Automatically extract and surface key assumptions from AI reasoning traces.
    \item Provide mechanisms for developers to validate, refine, or challenge these assumptions before acting on AI-generated recommendations.
    \item Leverage external knowledge sources, static analysis, or formal verification to assess the accuracy of AI assumptions.
\end{itemize}

By making AI assumptions explicit, developers can engage in informed decision-making rather than blindly following AI-generated suggestions. Work in this area could be based on related work from requirements engineering, for example~\cite{yang2018assumptions}.

\subsection{Multi-Model Alignment and Conflict Resolution}

As seen in our motivating examples, different AI models -- or even different runs of the same model -- may produce conflicting reasoning paths and conclusions. An interface should:
\begin{itemize}
    \item Compare reasoning outputs from multiple AI models, highlighting areas of agreement and disagreement.
    \item Provide confidence scores or uncertainty indicators to help developers assess which AI-generated insights are more reliable.
    \item Allow developers to actively engage with AI-generated conflicts, choosing reasoning paths based on their own expertise and external validation sources.
\end{itemize}

Rather than forcing developers to trust a single AI-generated thought process, the interface should facilitate a collaborative reasoning workflow, allowing human intuition and AI-generated insights to complement each other. Work in this area could be based on related work on inter-model agreement on software engineering tasks~\cite{ahmed2025can} and conflict resolution~\cite{dong2024revisiting}.

\subsection{Towards AI-Augmented Software Engineering}

Building such an interface requires rethinking how AI-generated reasoning integrates into software engineering workflows. In one extreme, traditional generative AI simply produces answers without exposing any reasoning, leaving developers without insight into its decision-making process. At the other extreme, new AI reasoning models generate extensive thought traces, often overwhelming users with verbose, unstructured reasoning. Neither extreme is ideal: developers need neither blind trust in opaque outputs nor the cognitive overload of sifting through every AI-generated assumption and inference.
What is needed is a balance: an AI that serves as a thought partner. Instead of merely dumping raw reasoning or providing unexplained answers, AI should present its thought process in a digestible manner.
% This vision calls for new research on human-AI interaction, focusing on how software engineers can efficiently interpret, validate, and apply AI-generated reasoning in real-world development scenarios.

\section{Research Roadmap and Open Challenges}

In the following, we outline research questions that should be addressed to integrate AI reasoning into software engineering workflows. 

\subsection{Understanding AI Reasoning in a Software Engineering Context}

The first two questions explore how AI-generated reasoning appears in software engineering tasks and domains.

\textit{RQ1. How can we collect and analyze large-scale corpora of AI-generated reasoning to understand its characteristics in software engineering?}
AI reasoning models generate complex, multi-step explanations, yet little is known about how these reasoning patterns manifest across different software engineering tasks. Empirical software engineering researchers should use these advancements in AI as an opportunity to ask new questions and analyze new data~\cite{treude2025generative}: They should systematically collect, categorize, and analyze AI-generated thought processes at scale to identify common reasoning structures, potential biases, and areas where AI reasoning aligns -- or conflicts -- with human decision-making. 

\textit{RQ2. How does AI-generated reasoning differ between different domains and tasks of software engineering?}
Software engineering spans diverse activities, from debugging to architectural design to performance optimization. AI reasoning might be more effective in certain domains (e.g., code generation) while struggling in others (e.g., long-term decisions). 
% Understanding the strengths and limitations of AI reasoning across different software engineering contexts requires systematic empirical studies that analyze variations in AI reasoning outputs across tasks.

\subsection{Summarization and Filtering of AI Reasoning}

The next two questions focus on extracting and presenting only the most relevant AI-generated insights.

\textit{RQ3. How can we automatically summarize AI reasoning traces to provide concise and relevant insights to developers?}
AI-generated thought traces are often too verbose, containing useful insights and irrelevant details. Developers need concise explanations that retain essential reasoning steps while filtering out noise. Research on natural language processing techniques, combined with empirical studies on developer information needs, could help refine summarization approaches tailored to software engineering tasks.

\textit{RQ4. How can we determine which parts of AI reasoning are most relevant to software engineers?}
Not all reasoning steps are equally useful, some provide critical insights, while others are redundant or misleading. Understanding how developers navigate AI-generated explanations requires investigating how they prioritize different types of reasoning when debugging, designing software architectures, or making design decisions. Studies on developer workflows and cognitive load could provide information on how reasoning should be structured for maximum usability.

\subsection{Validation and Assumption Checking}

The next two questions address identifying and verifying AI-generated assumptions.

\textit{RQ5. How can we extract and formalize implicit assumptions in AI-generated reasoning?}
AI reasoning models frequently make assumptions that shape their recommendations. If these assumptions are incorrect or misaligned with project requirements, developers may unknowingly accept flawed suggestions. Techniques from program analysis, specification mining, and knowledge representation could help detect and make these hidden assumptions explicit.

\textit{RQ6. How can we verify whether AI-generated assumptions align with software engineering best practices and are appropriate within the specific development context?}
While AI models generate reasoning based on training data, their assumptions do not always reflect best practices or align with the specific development context, including project requirements, tech stack, and constraints. Automatically comparing AI-generated insights against patterns from software repositories, verified guidelines, and project-specific factors could help assess whether AI recommendations are technically sound and practically applicable.

\subsection{Handling Conflicts and Divergent Reasoning Paths}

The next two questions examine resolving conflicting AI reasoning and handling uncertainty.

\textit{RQ7. How can we align and reconcile conflicting reasoning paths produced by different AI models?}
Different AI models -- or even different runs of the same model -- may produce divergent reasoning paths for the same problem. For example, one model may focus on performance, while another prioritizes maintainability. Understanding how to reconcile these perspectives is essential for integrating multiple AI reasoning sources into software development workflows. Conflict resolution strategies, such as multi-model consensus techniques or decision-support frameworks, could help address this issue. Trees of thought, similar to those used as a prompting technique~\cite{yao2024tree}, could be used to represent common and divergent reasoning paths.

\textit{RQ8. How can we quantify uncertainty in AI-generated reasoning and communicate it effectively to developers?}
AI reasoning models operate probabilistically, which means that they may not always reach the same conclusion given similar inputs. Developers need to understand the degree of confidence behind an AI-generated reasoning step. Exploring ways to represent uncertainty, whether through confidence scores, uncertainty visualization, or interactive explanations, could improve trust in AI-driven decision making.

\subsection{Human-AI Collaboration and Usability}

The final two questions explore how developers interact with and benefit from AI reasoning.

\textit{RQ9. How can we design interfaces that allow software engineers to interact efficiently with AI reasoning models?}
An AI-generated reasoning trace is only useful if developers can easily interpret and act upon it. Poorly designed interfaces could overwhelm users rather than help them. 
% Investigating human-computer interaction principles and developer experience research could provide insights into designing interfaces that balance detail with usability.

\textit{RQ10. How can we measure the impact of AI-generated reasoning on software engineering productivity and decision-making?}
Despite the promise of AI reasoning models, there is little empirical evidence on how they influence developer productivity and decision quality. Conducting controlled experiments and real-world case studies could help quantify their impact, identifying both benefits and potential drawbacks.

In closing, by addressing these questions, we can move towards an era where AI serves as an effective thought partner and enhances the ability of software engineers to reason about complex decisions, rather than overwhelm them with unstructured, difficult-to-interpret thought traces.

\section*{Acknowledgment}

We thank Cristian Cadar for inspiring some of the ideas presented in this paper during discussions at Greybox Coffee.

%%% -*-BibTeX-*-
%%% Do NOT edit. File created by BibTeX with style
%%% ACM-Reference-Format-Journals [18-Jan-2012].


\begin{thebibliography}{34}

%%% ====================================================================
%%% NOTE TO THE USER: you can override these defaults by providing
%%% customized versions of any of these macros before the \bibliography
%%% command.  Each of them MUST provide its own final punctuation,
%%% except for \shownote{} and \showURL{}.  The latter two
%%% do not use final punctuation, in order to avoid confusing it with
%%% the Web address.
%%%
%%% To suppress output of a particular field, define its macro to expand
%%% to an empty string, or better, \unskip, like this:
%%%
%%% \newcommand{\showURL}[1]{\unskip}   % LaTeX syntax
%%%
%%% \def \showURL #1{\unskip}           % plain TeX syntax
%%%
%%% ====================================================================

\ifx \showCODEN    \undefined \def \showCODEN     #1{\unskip}     \fi
\ifx \showISBNx    \undefined \def \showISBNx     #1{\unskip}     \fi
\ifx \showISBNxiii \undefined \def \showISBNxiii  #1{\unskip}     \fi
\ifx \showISSN     \undefined \def \showISSN      #1{\unskip}     \fi
\ifx \showLCCN     \undefined \def \showLCCN      #1{\unskip}     \fi
\ifx \shownote     \undefined \def \shownote      #1{#1}          \fi
\ifx \showarticletitle \undefined \def \showarticletitle #1{#1}   \fi
\ifx \showURL      \undefined \def \showURL       {\relax}        \fi
% The following commands are used for tagged output and should be
% invisible to TeX
\providecommand\bibfield[2]{#2}
\providecommand\bibinfo[2]{#2}
\providecommand\natexlab[1]{#1}
\providecommand\showeprint[2][]{arXiv:#2}

\bibitem[dee({[n.\,d.]})]%
        {deepmindGeminiFlash}
 \bibinfo{year}{[n.\,d.]}\natexlab{}.
\newblock \bibinfo{title}{{G}emini 2.0 {F}lash {T}hinking --- deepmind.google}.
\newblock \bibinfo{howpublished}{\url{https://deepmind.google/technologies/gemini/flash-thinking/}}.
\newblock
\newblock
\shownote{[Accessed 19-02-2025]}.


\bibitem[Ahmed et~al\mbox{.}(2025)]%
        {ahmed2025can}
\bibfield{author}{\bibinfo{person}{Toufique Ahmed}, \bibinfo{person}{Premkumar Devanbu}, \bibinfo{person}{Christoph Treude}, {and} \bibinfo{person}{Michael Pradel}.} \bibinfo{year}{2025}\natexlab{}.
\newblock \showarticletitle{Can LLMs replace manual annotation of software engineering artifacts?}. In \bibinfo{booktitle}{\emph{Proceedings of the International Conference on Mining Software Repositories}}.
\newblock


\bibitem[Ben-David et~al\mbox{.}(2015)]%
        {ben2015symbolic}
\bibfield{author}{\bibinfo{person}{Shoham Ben-David}, \bibinfo{person}{Baruch Sterin}, \bibinfo{person}{Joanne~M Atlee}, {and} \bibinfo{person}{Sandy Beidu}.} \bibinfo{year}{2015}\natexlab{}.
\newblock \showarticletitle{Symbolic model checking of product-line requirements using sat-based methods}. In \bibinfo{booktitle}{\emph{2015 IEEE/ACM 37th IEEE International Conference on Software Engineering}}, Vol.~\bibinfo{volume}{1}. IEEE, \bibinfo{pages}{189--199}.
\newblock


\bibitem[Cao et~al\mbox{.}(2024)]%
        {cao2024systematic}
\bibfield{author}{\bibinfo{person}{Sicong Cao}, \bibinfo{person}{Xiaobing Sun}, \bibinfo{person}{Ratnadira Widyasari}, \bibinfo{person}{David Lo}, \bibinfo{person}{Xiaoxue Wu}, \bibinfo{person}{Lili Bo}, \bibinfo{person}{Jiale Zhang}, \bibinfo{person}{Bin Li}, \bibinfo{person}{Wei Liu}, \bibinfo{person}{Di Wu}, {et~al\mbox{.}}} \bibinfo{year}{2024}\natexlab{}.
\newblock \showarticletitle{A systematic literature review on explainability for machine/deep learning-based software engineering research}.
\newblock \bibinfo{journal}{\emph{arXiv preprint arXiv:2401.14617}} (\bibinfo{year}{2024}).
\newblock


\bibitem[De~Moura and Bj{\o}rner(2010)]%
        {de2010bugs}
\bibfield{author}{\bibinfo{person}{Leonardo De~Moura} {and} \bibinfo{person}{Nikolaj Bj{\o}rner}.} \bibinfo{year}{2010}\natexlab{}.
\newblock \showarticletitle{Bugs, moles and skeletons: Symbolic reasoning for software development}. In \bibinfo{booktitle}{\emph{Automated Reasoning: 5th International Joint Conference, IJCAR 2010, Edinburgh, UK, July 16-19, 2010. Proceedings 5}}. Springer, \bibinfo{pages}{400--411}.
\newblock


\bibitem[Dehghani et~al\mbox{.}(2024)]%
        {dehghani2024trustworthy}
\bibfield{author}{\bibinfo{person}{Farzaneh Dehghani}, \bibinfo{person}{Mahsa Dibaji}, \bibinfo{person}{Fahim Anzum}, \bibinfo{person}{Lily Dey}, \bibinfo{person}{Alican Basdemir}, \bibinfo{person}{Sayeh Bayat}, \bibinfo{person}{Jean-Christophe Boucher}, \bibinfo{person}{Steve Drew}, \bibinfo{person}{Sarah~Elaine Eaton}, \bibinfo{person}{Richard Frayne}, {et~al\mbox{.}}} \bibinfo{year}{2024}\natexlab{}.
\newblock \showarticletitle{Trustworthy and Responsible AI for Human-Centric Autonomous Decision-Making Systems}.
\newblock \bibinfo{journal}{\emph{arXiv preprint arXiv:2408.15550}} (\bibinfo{year}{2024}).
\newblock


\bibitem[Dong et~al\mbox{.}(2024)]%
        {dong2024revisiting}
\bibfield{author}{\bibinfo{person}{Jinhao Dong}, \bibinfo{person}{Jun Sun}, \bibinfo{person}{Yun Lin}, \bibinfo{person}{Yedi Zhang}, \bibinfo{person}{Murong Ma}, \bibinfo{person}{Jin~Song Dong}, {and} \bibinfo{person}{Dan Hao}.} \bibinfo{year}{2024}\natexlab{}.
\newblock \showarticletitle{Revisiting the Conflict-Resolving Problem from a Semantic Perspective}. In \bibinfo{booktitle}{\emph{Proceedings of the 39th IEEE/ACM International Conference on Automated Software Engineering}}. \bibinfo{pages}{141--152}.
\newblock


\bibitem[Drozdal et~al\mbox{.}(2020)]%
        {drozdal2020trust}
\bibfield{author}{\bibinfo{person}{Jaimie Drozdal}, \bibinfo{person}{Justin Weisz}, \bibinfo{person}{Dakuo Wang}, \bibinfo{person}{Gaurav Dass}, \bibinfo{person}{Bingsheng Yao}, \bibinfo{person}{Changruo Zhao}, \bibinfo{person}{Michael Muller}, \bibinfo{person}{Lin Ju}, {and} \bibinfo{person}{Hui Su}.} \bibinfo{year}{2020}\natexlab{}.
\newblock \showarticletitle{Trust in AutoML: exploring information needs for establishing trust in automated machine learning systems}. In \bibinfo{booktitle}{\emph{Proceedings of the 25th international conference on intelligent user interfaces}}. \bibinfo{pages}{297--307}.
\newblock


\bibitem[Guo et~al\mbox{.}(2025)]%
        {guo2025deepseek}
\bibfield{author}{\bibinfo{person}{Daya Guo}, \bibinfo{person}{Dejian Yang}, \bibinfo{person}{Haowei Zhang}, \bibinfo{person}{Junxiao Song}, \bibinfo{person}{Ruoyu Zhang}, \bibinfo{person}{Runxin Xu}, \bibinfo{person}{Qihao Zhu}, \bibinfo{person}{Shirong Ma}, \bibinfo{person}{Peiyi Wang}, \bibinfo{person}{Xiao Bi}, {et~al\mbox{.}}} \bibinfo{year}{2025}\natexlab{}.
\newblock \showarticletitle{Deepseek-r1: Incentivizing reasoning capability in llms via reinforcement learning}.
\newblock \bibinfo{journal}{\emph{arXiv preprint arXiv:2501.12948}} (\bibinfo{year}{2025}).
\newblock


\bibitem[Haiduc et~al\mbox{.}(2010)]%
        {haiduc2010use}
\bibfield{author}{\bibinfo{person}{Sonia Haiduc}, \bibinfo{person}{Jairo Aponte}, \bibinfo{person}{Laura Moreno}, {and} \bibinfo{person}{Andrian Marcus}.} \bibinfo{year}{2010}\natexlab{}.
\newblock \showarticletitle{On the use of automated text summarization techniques for summarizing source code}. In \bibinfo{booktitle}{\emph{2010 17th Working conference on reverse engineering}}. IEEE, \bibinfo{pages}{35--44}.
\newblock


\bibitem[He et~al\mbox{.}(2025)]%
        {he2025llm}
\bibfield{author}{\bibinfo{person}{Junda He}, \bibinfo{person}{Christoph Treude}, {and} \bibinfo{person}{David Lo}.} \bibinfo{year}{2025}\natexlab{}.
\newblock \showarticletitle{LLM-Based Multi-Agent Systems for Software Engineering: Literature Review, Vision and the Road Ahead}.
\newblock \bibinfo{journal}{\emph{ACM Transactions on Software Engineering and Methodology}} (\bibinfo{year}{2025}).
\newblock


\bibitem[Jaech et~al\mbox{.}(2024)]%
        {jaech2024openai}
\bibfield{author}{\bibinfo{person}{Aaron Jaech}, \bibinfo{person}{Adam Kalai}, \bibinfo{person}{Adam Lerer}, \bibinfo{person}{Adam Richardson}, \bibinfo{person}{Ahmed El-Kishky}, \bibinfo{person}{Aiden Low}, \bibinfo{person}{Alec Helyar}, \bibinfo{person}{Aleksander Madry}, \bibinfo{person}{Alex Beutel}, \bibinfo{person}{Alex Carney}, {et~al\mbox{.}}} \bibinfo{year}{2024}\natexlab{}.
\newblock \showarticletitle{Openai o1 system card}.
\newblock \bibinfo{journal}{\emph{arXiv preprint arXiv:2412.16720}} (\bibinfo{year}{2024}).
\newblock


\bibitem[Kamran et~al\mbox{.}(2024)]%
        {kamran2024vision}
\bibfield{author}{\bibinfo{person}{Parnian Kamran}, \bibinfo{person}{Premkumar Devanbu}, {and} \bibinfo{person}{Caleb Stanford}.} \bibinfo{year}{2024}\natexlab{}.
\newblock \showarticletitle{Vision Paper: Proof-Carrying Code Completions}. In \bibinfo{booktitle}{\emph{Proceedings of the 39th IEEE/ACM International Conference on Automated Software Engineering Workshops}}. \bibinfo{pages}{35--42}.
\newblock


\bibitem[Kersten and Murphy(2006)]%
        {kersten2006using}
\bibfield{author}{\bibinfo{person}{Mik Kersten} {and} \bibinfo{person}{Gail~C Murphy}.} \bibinfo{year}{2006}\natexlab{}.
\newblock \showarticletitle{Using task context to improve programmer productivity}. In \bibinfo{booktitle}{\emph{Proceedings of the 14th ACM SIGSOFT international symposium on Foundations of software engineering}}. \bibinfo{pages}{1--11}.
\newblock


\bibitem[Li et~al\mbox{.}(2025)]%
        {li2025structured}
\bibfield{author}{\bibinfo{person}{Jia Li}, \bibinfo{person}{Ge Li}, \bibinfo{person}{Yongmin Li}, {and} \bibinfo{person}{Zhi Jin}.} \bibinfo{year}{2025}\natexlab{}.
\newblock \showarticletitle{Structured chain-of-thought prompting for code generation}.
\newblock \bibinfo{journal}{\emph{ACM Transactions on Software Engineering and Methodology}} \bibinfo{volume}{34}, \bibinfo{number}{2} (\bibinfo{year}{2025}), \bibinfo{pages}{1--23}.
\newblock


\bibitem[Liu et~al\mbox{.}(2024)]%
        {liu2024your}
\bibfield{author}{\bibinfo{person}{Jiawei Liu}, \bibinfo{person}{Chunqiu~Steven Xia}, \bibinfo{person}{Yuyao Wang}, {and} \bibinfo{person}{Lingming Zhang}.} \bibinfo{year}{2024}\natexlab{}.
\newblock \showarticletitle{Is your code generated by chatgpt really correct? rigorous evaluation of large language models for code generation}.
\newblock \bibinfo{journal}{\emph{Advances in Neural Information Processing Systems}}  \bibinfo{volume}{36} (\bibinfo{year}{2024}).
\newblock


\bibitem[Lu et~al\mbox{.}(2023)]%
        {lu2023llama}
\bibfield{author}{\bibinfo{person}{Junyi Lu}, \bibinfo{person}{Lei Yu}, \bibinfo{person}{Xiaojia Li}, \bibinfo{person}{Li Yang}, {and} \bibinfo{person}{Chun Zuo}.} \bibinfo{year}{2023}\natexlab{}.
\newblock \showarticletitle{LLaMA-Reviewer: Advancing code review automation with large language models through parameter-efficient fine-tuning}. In \bibinfo{booktitle}{\emph{2023 IEEE 34th International Symposium on Software Reliability Engineering (ISSRE)}}. IEEE, \bibinfo{pages}{647--658}.
\newblock


\bibitem[Mirzadeh et~al\mbox{.}(2024)]%
        {mirzadeh2024gsm}
\bibfield{author}{\bibinfo{person}{Iman Mirzadeh}, \bibinfo{person}{Keivan Alizadeh}, \bibinfo{person}{Hooman Shahrokhi}, \bibinfo{person}{Oncel Tuzel}, \bibinfo{person}{Samy Bengio}, {and} \bibinfo{person}{Mehrdad Farajtabar}.} \bibinfo{year}{2024}\natexlab{}.
\newblock \showarticletitle{Gsm-symbolic: Understanding the limitations of mathematical reasoning in large language models}.
\newblock \bibinfo{journal}{\emph{arXiv preprint arXiv:2410.05229}} (\bibinfo{year}{2024}).
\newblock


\bibitem[Sch{\"a}fer et~al\mbox{.}(2023)]%
        {schafer2023empirical}
\bibfield{author}{\bibinfo{person}{Max Sch{\"a}fer}, \bibinfo{person}{Sarah Nadi}, \bibinfo{person}{Aryaz Eghbali}, {and} \bibinfo{person}{Frank Tip}.} \bibinfo{year}{2023}\natexlab{}.
\newblock \showarticletitle{An empirical evaluation of using large language models for automated unit test generation}.
\newblock \bibinfo{journal}{\emph{IEEE Transactions on Software Engineering}} (\bibinfo{year}{2023}).
\newblock


\bibitem[Shi et~al\mbox{.}(2022)]%
        {shi2022buglistener}
\bibfield{author}{\bibinfo{person}{Lin Shi}, \bibinfo{person}{Fangwen Mu}, \bibinfo{person}{Yumin Zhang}, \bibinfo{person}{Ye Yang}, \bibinfo{person}{Junjie Chen}, \bibinfo{person}{Xiao Chen}, \bibinfo{person}{Hanzhi Jiang}, \bibinfo{person}{Ziyou Jiang}, {and} \bibinfo{person}{Qing Wang}.} \bibinfo{year}{2022}\natexlab{}.
\newblock \showarticletitle{Buglistener: identifying and synthesizing bug reports from collaborative live chats}. In \bibinfo{booktitle}{\emph{Proceedings of the 44th International Conference on Software Engineering}}. \bibinfo{pages}{299--311}.
\newblock


\bibitem[Tang et~al\mbox{.}(2023)]%
        {tang2023towards}
\bibfield{author}{\bibinfo{person}{Ziyi Tang}, \bibinfo{person}{Ruilin Wang}, \bibinfo{person}{Weixing Chen}, \bibinfo{person}{Keze Wang}, \bibinfo{person}{Yang Liu}, \bibinfo{person}{Tianshui Chen}, {and} \bibinfo{person}{Liang Lin}.} \bibinfo{year}{2023}\natexlab{}.
\newblock \showarticletitle{Towards causalgpt: A multi-agent approach for faithful knowledge reasoning via promoting causal consistency in llms}.
\newblock \bibinfo{journal}{\emph{arXiv preprint arXiv:2308.11914}} (\bibinfo{year}{2023}).
\newblock


\bibitem[Terragni et~al\mbox{.}(2025)]%
        {terragni2025future}
\bibfield{author}{\bibinfo{person}{Valerio Terragni}, \bibinfo{person}{Annie Vella}, \bibinfo{person}{Partha Roop}, {and} \bibinfo{person}{Kelly Blincoe}.} \bibinfo{year}{2025}\natexlab{}.
\newblock \showarticletitle{The Future of AI-Driven Software Engineering}.
\newblock \bibinfo{journal}{\emph{ACM Trans. Softw. Eng. Methodol.}} (\bibinfo{year}{2025}).
\newblock


\bibitem[Treude and Gerosa(2025)]%
        {treude2025developers}
\bibfield{author}{\bibinfo{person}{Christoph Treude} {and} \bibinfo{person}{Marco~A Gerosa}.} \bibinfo{year}{2025}\natexlab{}.
\newblock \showarticletitle{How Developers Interact with AI: A Taxonomy of Human-AI Collaboration in Software Engineering}. In \bibinfo{booktitle}{\emph{2nd ACM International Conference on AI Foundation Models and Software Engineering}}.
\newblock


\bibitem[Treude and Storey(2025)]%
        {treude2025generative}
\bibfield{author}{\bibinfo{person}{Christoph Treude} {and} \bibinfo{person}{Margaret-Anne Storey}.} \bibinfo{year}{2025}\natexlab{}.
\newblock \showarticletitle{Generative AI and Empirical Software Engineering: A Paradigm Shift}.
\newblock \bibinfo{journal}{\emph{arXiv preprint arXiv:2502.08108}} (\bibinfo{year}{2025}).
\newblock


\bibitem[Ullah et~al\mbox{.}(2024)]%
        {ullah2024llms}
\bibfield{author}{\bibinfo{person}{Saad Ullah}, \bibinfo{person}{Mingji Han}, \bibinfo{person}{Saurabh Pujar}, \bibinfo{person}{Hammond Pearce}, \bibinfo{person}{Ayse Coskun}, {and} \bibinfo{person}{Gianluca Stringhini}.} \bibinfo{year}{2024}\natexlab{}.
\newblock \showarticletitle{{LLMs} Cannot Reliably Identify and Reason About Security Vulnerabilities (Yet?): A Comprehensive Evaluation, Framework, and Benchmarks}. In \bibinfo{booktitle}{\emph{Proceedings of the IEEE Symposium on Security and Privacy}}.
\newblock


\bibitem[Wang et~al\mbox{.}({[n.\,d.]})]%
        {wangunlocking}
\bibfield{author}{\bibinfo{person}{Guoyin Wang}, \bibinfo{person}{Shengyu Zhang}, \bibinfo{person}{Tianyu Zhan}, \bibinfo{person}{Zhouzhou Shen}, \bibinfo{person}{Jiwei Li}, \bibinfo{person}{Xueyu Hu}, \bibinfo{person}{Xiaofei Sun}, \bibinfo{person}{Fei Wu}, \bibinfo{person}{Gelei Deng}, \bibinfo{person}{Jie Zhang}, {et~al\mbox{.}}} \bibinfo{year}{[n.\,d.]}\natexlab{}.
\newblock \showarticletitle{Unlocking the Mysteries of OpenAI o1: A Survey of the Reasoning Abilities of Large Language Models}.
\newblock  (\bibinfo{year}{[n.\,d.]}).
\newblock


\bibitem[Wang et~al\mbox{.}(2024)]%
        {wang2024investigating}
\bibfield{author}{\bibinfo{person}{Ruotong Wang}, \bibinfo{person}{Ruijia Cheng}, \bibinfo{person}{Denae Ford}, {and} \bibinfo{person}{Thomas Zimmermann}.} \bibinfo{year}{2024}\natexlab{}.
\newblock \showarticletitle{Investigating and designing for trust in ai-powered code generation tools}. In \bibinfo{booktitle}{\emph{The 2024 ACM Conference on Fairness, Accountability, and Transparency}}. \bibinfo{pages}{1475--1493}.
\newblock


\bibitem[Wei et~al\mbox{.}(2022)]%
        {wei2022chain}
\bibfield{author}{\bibinfo{person}{Jason Wei}, \bibinfo{person}{Xuezhi Wang}, \bibinfo{person}{Dale Schuurmans}, \bibinfo{person}{Maarten Bosma}, \bibinfo{person}{Fei Xia}, \bibinfo{person}{Ed Chi}, \bibinfo{person}{Quoc~V Le}, \bibinfo{person}{Denny Zhou}, {et~al\mbox{.}}} \bibinfo{year}{2022}\natexlab{}.
\newblock \showarticletitle{Chain-of-thought prompting elicits reasoning in large language models}.
\newblock \bibinfo{journal}{\emph{Advances in neural information processing systems}}  \bibinfo{volume}{35} (\bibinfo{year}{2022}), \bibinfo{pages}{24824--24837}.
\newblock


\bibitem[Weng et~al\mbox{.}(2023)]%
        {weng2023mastering}
\bibfield{author}{\bibinfo{person}{Yixuan Weng}, \bibinfo{person}{Minjun Zhu}, \bibinfo{person}{Fei Xia}, \bibinfo{person}{Bin Li}, \bibinfo{person}{Shizhu He}, \bibinfo{person}{Kang Liu}, {and} \bibinfo{person}{Jun Zhao}.} \bibinfo{year}{2023}\natexlab{}.
\newblock \showarticletitle{Mastering Symbolic Operations: Augmenting Language Models with Compiled Neural Networks}.
\newblock \bibinfo{journal}{\emph{arXiv preprint arXiv:2304.01665}} (\bibinfo{year}{2023}).
\newblock


\bibitem[Yang et~al\mbox{.}(2018)]%
        {yang2018assumptions}
\bibfield{author}{\bibinfo{person}{Chen Yang}, \bibinfo{person}{Peng Liang}, {and} \bibinfo{person}{Paris Avgeriou}.} \bibinfo{year}{2018}\natexlab{}.
\newblock \showarticletitle{Assumptions and their management in software development: A systematic mapping study}.
\newblock \bibinfo{journal}{\emph{Information and Software Technology}}  \bibinfo{volume}{94} (\bibinfo{year}{2018}), \bibinfo{pages}{82--110}.
\newblock


\bibitem[Yao et~al\mbox{.}(2024)]%
        {yao2024tree}
\bibfield{author}{\bibinfo{person}{Shunyu Yao}, \bibinfo{person}{Dian Yu}, \bibinfo{person}{Jeffrey Zhao}, \bibinfo{person}{Izhak Shafran}, \bibinfo{person}{Tom Griffiths}, \bibinfo{person}{Yuan Cao}, {and} \bibinfo{person}{Karthik Narasimhan}.} \bibinfo{year}{2024}\natexlab{}.
\newblock \showarticletitle{Tree of thoughts: Deliberate problem solving with large language models}.
\newblock \bibinfo{journal}{\emph{Advances in Neural Information Processing Systems}}  \bibinfo{volume}{36} (\bibinfo{year}{2024}).
\newblock


\bibitem[Yu et~al\mbox{.}(2018)]%
        {yu2018symbolic}
\bibfield{author}{\bibinfo{person}{Hengbiao Yu}, \bibinfo{person}{Zhenbang Chen}, \bibinfo{person}{Ji Wang}, \bibinfo{person}{Zhendong Su}, {and} \bibinfo{person}{Wei Dong}.} \bibinfo{year}{2018}\natexlab{}.
\newblock \showarticletitle{Symbolic verification of regular properties}. In \bibinfo{booktitle}{\emph{Proceedings of the 40th International Conference on Software Engineering}}. \bibinfo{pages}{871--881}.
\newblock


\bibitem[Zhang et~al\mbox{.}(2024)]%
        {zhang2024automatic}
\bibfield{author}{\bibinfo{person}{Yuxia Zhang}, \bibinfo{person}{Zhiqing Qiu}, \bibinfo{person}{Klaas-Jan Stol}, \bibinfo{person}{Wenhui Zhu}, \bibinfo{person}{Jiaxin Zhu}, \bibinfo{person}{Yingchen Tian}, {and} \bibinfo{person}{Hui Liu}.} \bibinfo{year}{2024}\natexlab{}.
\newblock \showarticletitle{Automatic commit message generation: A critical review and directions for future work}.
\newblock \bibinfo{journal}{\emph{IEEE Transactions on Software Engineering}} (\bibinfo{year}{2024}).
\newblock


\bibitem[Zhong and Wang(2024)]%
        {zhong2024can}
\bibfield{author}{\bibinfo{person}{Li Zhong} {and} \bibinfo{person}{Zilong Wang}.} \bibinfo{year}{2024}\natexlab{}.
\newblock \showarticletitle{Can LLM Replace Stack Overflow? A Study on Robustness and Reliability of Large Language Model Code Generation}. In \bibinfo{booktitle}{\emph{Proceedings of the AAAI Conference on Artificial Intelligence}}, Vol.~\bibinfo{volume}{38}. \bibinfo{pages}{21841--21849}.
\newblock


\end{thebibliography}
\end{document}